\documentclass[twoside]{article}
\usepackage{qic}

\newcommand{\ket}[1]{\vert #1 \rangle}

\begin{document}
\setlength{\textheight}{8.0truein}    
\runninghead{Security of quantum key distribution using weak coherent states with nonrandom phases }
            {H.-K. Lo and J. Preskill }

\normalsize\textlineskip
\thispagestyle{empty}
\setcounter{page}{1}


\vspace*{0.88truein}

\alphfootnote

\fpage{1}

\centerline{\bf
SECURITY OF QUANTUM KEY DISTRIBUTION USING}
\vspace*{0.035truein}
\centerline{\bf WEAK COHERENT STATES WITH NONRANDOM PHASES}
\vspace*{0.37truein}
\centerline{\footnotesize
HOI-KWONG LO}
\vspace*{0.015truein}
\centerline{\footnotesize\it Center for Quantum Information and Quantum Control }
\baselineskip=10pt
\centerline{\footnotesize\it Department of Electrical and Computer Engineering and Department of Physics}
\baselineskip=10pt
\centerline{\footnotesize\it University of Toronto, Toronto, Canada M5G 3G4}
\vspace*{10pt}
\centerline{\footnotesize 
JOHN PRESKILL}
\vspace*{0.015truein}
\centerline{\footnotesize\it Institute for Quantum Information}
\baselineskip=10pt
\centerline{\footnotesize\it California Institute of Technology,  Pasadena, CA 91125, USA}
\vspace*{0.225truein}

%
\vspace*{0.21truein}
%
\abstracts{
We prove the security of the Bennett-Brassard (BB84) quantum key distribution protocol in the case where the key information is encoded in the relative phase of a coherent-state reference pulse and a weak coherent-state signal pulse, as in some practical implementations of the protocol. In contrast to previous work, our proof applies even if the eavesdropper knows the phase of the reference pulse, provided that this phase is not modulated by the source, and even if the reference pulse is bright. The proof also applies to the case where the key is encoded in the photon polarization of a weak coherent-state pulse with a known phase, but only if the phases of the four BB84 signal states are judiciously chosen. The achievable key generation rate scales quadratically with the transmission in the channel, just as for BB84 with phase-randomized weak coherent-state signals (when decoy states are not used). For the case where the phase of the reference pulse {\em is} strongly modulated by the source, we exhibit an explicit attack that allows the eavesdropper to learn every key bit in a parameter regime where a protocol using phase-randomized signals is provably secure. 
}{}{}
\vspace*{10pt}

\vspace*{1pt}\textlineskip    

\section{Introduction}

In quantum key distribution (QKD) \cite{BB84}, two parties (Alice and Bob) use quantum signals to establish a shared key that can be used to encrypt and decrypt classical messages. An eavesdropper (Eve) who collects information about the key by interacting with the signals produces a detectable disturbance; therefore Alice and Bob can detect the eavesdropper's activity, and they can reject the key if they fear that the eavesdropper knows too much about it. But if the detected disturbance is weak enough, then Alice and Bob can use classical error correction and privacy amplification protocols to extract a shared key that is very nearly uniformly distributed and almost certainly private \cite{mayers,lo-chau,biham,ShorPreskill,koashi-preskill,christandl-renner-ekert,renner-gisin-kraus}. The security of the QKD protocol is said to be {\em unconditional}, because the security can be proven for any attack consistent with the laws of quantum physics, and without any assumptions about computational hardness.

Experiments have recently demonstrated the feasibility of QKD over 150 km telecom fibers \cite{NEC,GYS}, and at least two firms are now marketing commercial QKD systems \cite{commercial}. But how secure are these systems, really? To assess the security of practical implementations of QKD, it is important to consider how well the actual systems match the performance assumed in the security proofs. In particular, the signals used in typical practical realizations of QKD are dim laser pulses, which occasionally contain more than one photon. Multi-photon signals together with loss in the optical fiber can threaten security, but proofs of security for QKD using weak coherent states have been found \cite{ilm,GLLP}. (We note that for QKD protocols that use decoy states \cite{Hwang,Decoy,Wang}, security can be proven even for rather strong coherent-state signals. In this paper, however, we will focus on QKD with weak coherent states.)

A key assumption in the security proofs in \cite{ilm,GLLP} (and also in \cite{Decoy}) is that the phase of the quantum signal is uniformly random. A coherent state of one mode of the electromagnetic field can be expressed as
\begin{equation}
|\alpha\rangle = e^{-|\alpha|^2/2}\sum_{n=0}^\infty \frac{\alpha^n}{\sqrt{n!}} |n\rangle~,
\end{equation}
where $|n\rangle$ denotes the state with photon number $n$. We may write $\alpha=\sqrt{\nu}e^{i\theta}$, where $\nu=|\alpha|^2$ denotes the mean photon number and $e^{i\theta}$ is the {\em phase} of the coherent state. To an eavesdropper with no {\it a priori} knowledge of the phase, a signal whose phase is selected uniformly at random is indistinguishable from the state
\begin{equation}
\rho_\nu = \int_0^{2\pi} \frac{d\theta}{2\pi} |\sqrt{\nu}e^{i\theta}\rangle\langle 
\sqrt{\nu}e^{i\theta}|=e^{-\nu}\sum_{n=0}^\infty  \frac{\nu^n}{n!} |n\rangle\langle n|~,
\end{equation}
a Poisson distributed mixture of photon number eigenstates. Therefore, for a security analysis, we may suppose that a source emitting weak coherent state signals is actually emitting signals in the state $\rho_\nu$.

With probability $p_0=e^{-\nu}$, which is close to one for small $\nu$, the source emits no photon; exactly one photon is emitted with probability $p_1=\nu e^{-\nu}$. The probability that two or more photons are emitted is
\begin{equation}
p_M= 1- e^{-\nu}\left(1+\nu\right))\le \frac{1}{2} \nu^2~.
\end{equation}
Multi-photons can pose a security risk, but if each signal has a random phase, $p_M$ is sufficiently small, and the loss in the channel is not too high, then it is possible to prove security of the Bennett-Brassard (BB84) protocol \cite{BB84} against arbitrary eavesdropping attacks \cite{ilm,GLLP}.

However, no previously known security proof applies if the eavesdropper has some {\em a priori} knowledge about the phase of the signal states. Conceivably, such phase information might be accessible in realistic implementations of QKD. For example, in a ``plug-and-play'' scheme \cite{plug-and-play}, a strong signal is sent from Bob to Alice, who attenuates and modulates the signal before returning it to Bob; in unidirectional schemes as well, strong ancillary pulses are sometimes used to monitor the channel. The phase of a strong pulse is accurately measurable in principle, and could be correlated with the phase of the key-generating pulse. Furthermore, the key information itself might be encoded in the relative phase of a bright reference pulse and a weak signal pulse, in which case the eavesdropper could plausibly learn the phase of the reference pulse. Even if strong pulses are not used, the phase coherence of a realistic source might be maintained during the emission of many weak signals, allowing the phase to be determined accurately. 

For all these reasons, it is worthwhile to investigate the security of QKD under the assumption that the eavesdropper knows something about the phase of the signals. Our main result in this paper is a proof of security for the BB84 protocol in the case where the key information is encoded in the relative phase of a reference pulse and a signal pulse. Our proof works even if the reference pulse is bright, with a phase known by the eavesdropper. It also applies if the key is encoded in the photon polarization of weak coherent states with nonrandom phases, provided the phases of the signal states are chosen appropriately. The proof is founded on the observation that, if the signal pulse is weak, Alice's source reveals relatively little information to Eve about the basis that Alice uses to encode her key bits. Privacy of the final key is demonstrated using an argument due to Koashi \cite{koashi,koashi-realistic} that invokes the uncertainty principle. We also point out that this argument establishes that the key distribution protocol is {\em universally composable} --- the key can be used in any subsequent application without compromising security. 

We describe our quasi-realistic model for sources and detectors in Sec.~\ref{sec:modeling}, and present the security analysis (which relies heavily on Koashi's ideas \cite{koashi,koashi-realistic}) in Sec.~\ref{sec:security}, Sec.~\ref{sec:bright}, and Appendix A. In Sec.~\ref{sec:comparison} we note that the key generation rate when the signals have nonrandom phases is comparable to the rate for phase randomized signals (when decoy states are not used). In Sec.~\ref{sec:ukd} and Appendix B we point out that if the key is encoded in the photon polarization, and the (nonrandom) phases of the signals are not chosen judiciously, then BB84 is vulnerable to an {\em unambiguous key discrimination} attack, a measurement that, when conclusive, determines the key bit with certainty. Sec.~\ref{sec:conclusion} contains our conclusions. 

\section{Modeling sources and detectors}
\label{sec:modeling}
\subsection{The source}
In the ideal BB84 protocol \cite{BB84}, each signal is carried by a single qubit sent by Alice and received by Bob. The qubit encodes a key bit in one of two conjugate orthonormal bases, which we will call the $x$ basis and the $y$ basis. When Alice uses the $x$ basis, her signal states are
\begin{eqnarray}
|0_X\rangle&=& \left(|0_Z\rangle + |1_Z\rangle\right)/\sqrt{2}~,\nonumber\\
 |1_X\rangle&=& \left(|0_Z\rangle -|1_Z\rangle\right)/\sqrt{2}~,
\end{eqnarray}
the eigenstates with eigenvalues $\pm 1$ of the Pauli operator
\begin{equation}
X=
\pmatrix{0&1 \cr 1&0}
~;
\end{equation}
when Alice uses the $y$ basis, her signal states are
\begin{eqnarray}
|0_Y\rangle&=& \left(|0_Z\rangle + i|1_Z\rangle\right)/\sqrt{2}~,\nonumber\\
 |1_Y\rangle&=& \left(|0_Z\rangle -i|1_Z\rangle\right)/\sqrt{2}~,
\end{eqnarray}
the eigenstates with eigenvalues $\pm 1$ of the Pauli operator
\begin{equation}
Y=
\pmatrix{
0&-i\cr i&0
}~.
\end{equation}
Alice's source emits one of these four states, chosen equiprobably. Bob measures the qubit that he receives in either the $x$ basis or the $y$ basis, chosen equiprobably, to determine his key bit. Later, through public discussion, Alice and Bob ``sift'' their key by retaining only the key bits for which Bob measured in the same basis that Alice used. The final key is extracted from the sifted key via a classical protocol that reconciles Alice's key with Bob's and amplifies the privacy of the key. 

In practice, the key information is carried by dim laser pulses transmitted through an optical fiber. In the protocol we will analyze, Alice's source emits one of four states of a pair of photon modes (described by annihilation operators $a_R$ and $a_S$):
\begin{eqnarray} \label{bb84states}
\ket{\tilde 0_X} &=& e^{-|\alpha|^2}e^{\alpha\left(a_R^\dagger+a_S^\dagger\right)}\ket{\rm vac}_R\otimes \ket{\rm vac}_S=\ket{\alpha}_R\otimes \ket{\alpha}_S~, \nonumber \\
\ket{\tilde 1_X} &=& e^{-|\alpha|^2}e^{\alpha\left(a_R^\dagger-a_S^\dagger\right)}\ket{\rm vac}_R\otimes \ket{\rm vac}_S=\ket{\alpha}_R\otimes \ket{-\alpha}_S~, \nonumber \\
\ket{\tilde 0_Y} &=& e^{-|\alpha|^2}e^{\alpha\left(a_R^\dagger+ia_S^\dagger\right)}\ket{\rm vac}_R\otimes \ket{\rm vac}_S=\ket{\alpha}_R\otimes \ket{i\alpha}_S~, \nonumber \\
\ket{\tilde 0_Y} &=& e^{-|\alpha|^2}e^{\alpha\left(a_R^\dagger-ia_S^\dagger\right)}\ket{\rm vac}_R\otimes \ket{\rm vac}_S=\ket{\alpha}_R\otimes \ket{-i\alpha}_S~. \nonumber\\
\end{eqnarray}
Here the phase of $\alpha$ is defined relative to a fixed classical phase reference frame that Eve can access. We will refer to the state of mode $S$ as the ``signal'' pulse and to the state of mode $R$ as the ``reference'' pulse. The key is encoded in the relative phase of the two pulses, and it is actually quite important for our analysis that the phase of the reference pulse is the same in all four of the signal states --- {\em only} the phase of the signal pulse is modulated by the source. (We will discuss in Sec.~\ref{sec:ukd} and Appendix \ref{app:ukd} how security can be compromised when the phase of the reference pulse is also modulated.) In some implementations, the two pulses are spatially separated in the optical fiber by a distance large compared to the pulse width; this scheme is called ``phase encoding.'' In some implementations, the two modes represent the polarization states of the same spatial mode; this scheme is called polarization encoding. Phase encoding has been used in most fiber-based QKD experiments. Our security analysis will also apply to phase encoding in the case where the reference pulse is brighter than the signal pulse. We will return to this generalization in Sec.~\ref{sec:bright}, but for now we will stick with the signals eq.~(\ref{bb84states}), which could arise in either a phase-encoding or polarization-encoding scheme.

To summarize, the crucial features of our source model are:  (1) the {\em single-mode signal pulse assumption} (the signal pulse is carried by a single bosonic mode, the same mode for all four of the signal states), and  (2) the {\em unmodulated reference pulse assumption} (the state of the reference pulse is the same for all four signal states). These assumptions are reasonably well fulfilled by realistic sources used for the BB84 protocol with phase-encoded signals. Previous proofs of the security of BB84 have used ($2'$) the {\em phase-randomization assumption} (the phase of the reference pulse, but not of course the relative phase of signal and reference pulse that encodes the key information, is chosen uniformly at random and is not known by the eavesdropper). Our main new contribution in this paper is to prove security of BB84 using assumption (2) instead of assumption ($2'$). 

\subsection{The detectors}
When Bob wants to measure the signal he receives in the $x$ basis, he (in effect) combines the two modes in an interferometer, and directs modes $a_{0,X}$, $a_{1,X}$ to two different ``threshold'' photon detectors, where
\begin{eqnarray} 
\label{bobs-x-measurement}
a_{0,X} &=& \left(a_R+a_S\right)/\sqrt{2}~,\nonumber\\
a_{1,X} &=& \left(a_R-a_S\right)/\sqrt{2}~.
\end{eqnarray}
Therefore, if Alice sends $|\tilde 0_X\rangle$ and the ideal (unattenuated) signal enters Bob's measurement device (which is configured for $x$ measurement), then detector 0 receives a coherent state with mean photon number 
\begin{equation}
\mu=2|\alpha|^2
\end{equation}
and detector 1 receives the vacuum state. Likewise, if Alice sends $|\tilde 1_X\rangle$, then detector 0 receives the vacuum, and detector 1 receives the coherent state with mean photon number $\mu$. 
When Bob wants to measure in the $y$ basis, the modes directed to the photon detectors are
\begin{eqnarray} 
\label{bobs-y-measurement}
a_{0,Y} &=& \left(a_R+ia_S\right)/\sqrt{2}~,\nonumber\\
a_{1,Y} &=& \left(a_R-ia_S\right)/\sqrt{2}~;
\end{eqnarray}
again, depending on the key bit, one mode receives mean photon number $\mu$ while the other receives vacuum.

We assume that each detector used by Bob to measure the signal he receives is a ``threshold'' photon detector. This means that the detector cannot distinguish a single photon from many photons. An {\em ideal} threshold detector ``clicks'' (registers a count) if it collects one or more photons, and does not click if it receives no photon; thus it performs a POVM with the two outcomes
\begin{eqnarray}
&&E_{\rm no ~ click}= |{\rm vac}\rangle\langle {\rm vac}|~,\nonumber\\
&&E_{\rm click}= I-|{\rm vac}\rangle\langle {\rm vac}|~.
\end{eqnarray}
But in this paper we consider a threshold detector that is not ideal: we allow the detector to have imperfect efficiency (it sometimes fails to click even when it receives one or more photons) and to record occasional ``dark counts'' (it sometimes clicks even when it receives the vacuum state). 

Because Bob uses two threshold detectors, there are four possible outcomes when he measures a signal.
If neither detector clicks, then the measurement is inconclusive and the signal is rejected. If detector $b\in \{0,1\}$ clicks and the other detector does not, then Bob records the key bit $b$. If {\em both} detectors click, then Bob records a key bit that he chooses uniformly at random. As noted in \cite{ilm} and as we will explain below, it is important for the security analysis that double clicks are interpreted as key bits rather than regarded as inconclusive measurements. Actually, the security proof works the same way no matter how we map the double click events to key bits, but recording a random key bit is a particularly symmetric and natural choice. Given any fixed prescription for mapping double clicks to key bits, Bob's measurement realizes a POVM with three outcomes: 0, 1, and (when neither detector clicks) inconclusive.

Thus the protocol prescribes that Bob perform one of two possible POVMs, depending on whether Bob declares the $x$ basis or the $y$ basis, where each POVM has three outcomes. In fact, for our security proof we will not need to specify Bob's measurement in detail. All we require is that the measurement satisfy one assumption: Bob's three-outcome POVM is equivalent to a (basis-dependent) two-outcome POVM that determines the key bit, preceded by a {\em basis-independent} ``filter'' that allows the signal to pass if and only if Bob's measurement conclusively determines the key bit.

Following Koashi \cite{koashi-realistic}, we note that this property, that the inconclusive measurement outcome can be attributed to a basis-independent filter, follows if we adopt a particular well-motivated model of dark counts and detector inefficiency, and if we suppose that both of Bob's detectors have the same efficiency. First we note that the inconclusive outcome can occur only if neither detector has a dark count. Let us model the dark counts as a random background process, where the rate of background counts is an intrinsic property of the detector, not dependent on the state of the mode that is being measured. That is, a background click occurs in detector 0 with probability $d_0$, whether mode $a_{0,X}$ or mode $a_{0,Y}$ is being measured, and in detector 1 with probability $d_1$, whether mode $a_{1,X}$ or mode $a_{1,Y}$ is being measured. Thus, each time a signal is received, the probability $(1-d_0)(1-d_1)$ that there is no background click in either detector does not depend on whether Bob declares the $x$ basis or the $y$ basis. 

Furthermore, let us also assume \cite{walls-milburn} that when mode $a_{0,X}$ (or $a_{0,Y}$) contains $n$ photons the probability that detector 0 fails to click is 
\begin{equation}
\label{no-click-prob}
{\rm Prob}(n~{\rm photons, no~click})= (1-\xi)^n~,
\end{equation}
where $\xi$ is the detector's efficiency, and similarly for detector 1. We may imagine that the detector induces decoherence in the photon number measurement before measuring --- the real detector might not actually destroy the coherence of a superposition of states with two different nonzero photon numbers, but we are entitled to suppose that it does, because if so Eve's information and the measurement outcome would not be affected. Thus, each signal Bob receives can be regarded as a mixture of eigenstates of the total photon number $n$, where $m \le n$ of the photons are directed to detector 0 and the remaining $n-m$ photons are directed to mode 1. But if both detectors have the same efficiency $\xi$, and if the probability of no click is given by eq.~(\ref{no-click-prob}), then the probability that {\em neither} detector clicks depends only on $n$, not on $m$, and therefore it makes no difference whether Bob's detectors receive modes $a_{0,X}$ and  $a_{1,X}$ or $a_{0,Y}$ and $a_{1,Y}$; in other words the probability of an inconclusive outcome does not depend on whether Bob measures in the $x$ basis or the $y$ basis.

We conclude, then, that when Bob receives an $n$ photon signal, the probability that no click occurs in either detector, so that the measurement is inconclusive, is
\begin{equation}
{\rm Prob(}n,{\rm inconclusive)}= (1-d_0)(1-d_1)(1-\xi)^n~,
\end{equation}
which does not depend on whether Bob declares the $x$ basis or the $y$ basis. Thus, we can imagine that a basis-independent filter is applied that measures the total photon number $n$ and blocks the signal with probability ${\rm Prob(}n,{\rm inconclusive)}$; if the signal passes the filter, Bob then performs a two-outcome POVM, either $M_x$ or $M_y$ depending on whether Bob declares the $x$ basis or the $y$ basis. 

In order to describe Bob's measurement this way, we need a prescription for mapping the double click events to key bits. This prescription is needed because the probability of a double click event might depend on Bob's declared basis; therefore if the double clicks were regarded as inconclusive we would not be able to attribute the inconclusive outcome to a basis-independent filter. The prescription is incorporated into the definition of the measurements $M_x$ and $M_y$, and for the security proof, the details of this prescription do not matter.

We note that if Bob receives a coherent state with mean photon number $\mu$, then the inconclusive outcome occurs with probability 
\begin{eqnarray}
\label{coherent-inconclusive}
{\rm Prob(}\mu,{\rm inconclusive)}&=& e^{-\mu}\sum_{n=0}^\infty \frac{\mu^n}{n!}{\rm Prob(}n,{\rm inconclusive)}\nonumber\\
 &=&(1-d_0)(1-d_1)e^{-\xi\mu}~.
\end{eqnarray}
In particular, eq.~(\ref{coherent-inconclusive}) applies if Bob receives one of the signal states in eq.~(\ref{bb84states}) and performs the measurement specified in eq.~(\ref{bobs-x-measurement}) or eq.~(\ref{bobs-y-measurement}), where $\mu=2|\alpha|^2$. In that case, when Alice and Bob use the same basis, double clicks occur only because of dark counts, and when the measurement is conclusive and there are no dark counts Bob's key bit always agrees with Alice's. Attenuation in the channel reduces the probability of a conclusive outcome, but does not cause bit errors if the signals are otherwise unmodified and if Bob's measurement is ideal. 

To summarize, the crucial feature of our detector model is the {\em basis-independent filter assumption}, which says that Bob's measurement can be modeled by a basis-independent filter that removes all the signals for which the measurement of the key bit is inconclusive, followed by a (basis-dependent) two-outcome POVM that always produces a conclusive result. This assumption is reasonably fulfilled by realistic measurement devices used in BB84 provided that (1) Bob uses two threshold detectors with the same efficiency, (2) the probability that the detector fails to click decays exponentially as a function of the number of photons received (as in the standard theory of photodetection), and (3) a prescription (either deterministic or probabilistic) is adopted for mapping double click events to key bits. We note that attacks have been proposed and analyzed \cite{mismatch1,mismatch2} that exploit a mismatch in the efficiency of Bob's two photon detectors, highlighting the importance of assumption (1).

The basis-independent filter assumption limits Eve's ability to enhance her information about the signals by taking advantage of detector inefficiency. But we emphasize that our model still incorporates quite general flaws in the detection system (including for example a reduction in visibility due to an imperfect alignment of Bob's interferometer), because the POVMs $M_x$ and $M_y$ are arbitrary. Security is not compromised even if the detectors are very noisy, because in that case the bit error rate will be high and the key will almost certainly be rejected.

\section{Security analysis}
\label{sec:security}

\subsection{The unbalanced quantum coin}

We can analyze the security of this protocol by applying the methods of \cite{GLLP,koashi}, where security was proven for the case where Alice's source emits signals that, averaged over the key bits, have a small dependence on the basis. To apply this method, we must quantify the basis-dependence of the signal set eq.~(\ref{bb84states}).

It is convenient to imagine that Alice launches each signal by performing a perfect measurement on a qubit that is entangled with the signal mode (the reference mode is always prepared in the state $|\alpha\rangle$ no matter which signal is being sent). When Alice wants to send one of the two signals $\{|\tilde 0_X\rangle, |\tilde 1_X\rangle\}$, we choose the entangled state of her qubit and the signal mode to be
\begin{equation}
|\Psi_x\rangle = \left(|0_X\rangle \otimes |\alpha\rangle + |1_X\rangle \otimes |-\alpha\rangle\right)/\sqrt{2}~,
\end{equation}
and we instruct Alice to determine her key bit by measuring her qubit in the basis $\{|0_X\rangle,|1_X\rangle\}$. When Alice wants to send one of the two signals $\{|\tilde 0_Y\rangle, |\tilde 1_Y\rangle\}$, we choose the entangled state of her qubit and the mode to be
\begin{equation}
\label{Psi-y}
|\Psi_y\rangle = \left(|1_Y\rangle \otimes |i\alpha\rangle + |0_Y\rangle \otimes |-i\alpha\rangle\right)/\sqrt{2}~,
\end{equation}
and we instruct Alice to determine her key bit by measuring her qubit in the basis $\{|1_Y\rangle,|0_Y\rangle\}$. See Fig.~\ref{fig:setup}.

Now we notice that the states $|\Psi_x\rangle$ and $|\Psi_y\rangle$ are hard to distinguish when $\alpha$ is small. 
Using
\begin{eqnarray}
&&\langle i\alpha|\alpha\rangle= e^{-|\alpha|^2} e^{{-}i|\alpha|^2}=\langle {-}i\alpha|{-}\alpha\rangle~,\nonumber\\
&&\langle {-}i\alpha|\alpha\rangle= e^{-|\alpha|^2} e^{i|\alpha|^2}=\langle i\alpha|{-}\alpha\rangle~,
\end{eqnarray}
we find
\begin{eqnarray}
\langle \Psi_y|\Psi_x\rangle &=& e^{-|\alpha|^2}\left(\cos|\alpha|^2 + \sin |\alpha|^2\right)\nonumber\\
&=& e^{-\mu/2}\left(\cos(\mu/2) + \sin (\mu/2)\right)\nonumber\\
&=& 1-\mu^2/4 +O(\mu^3)~.
\end{eqnarray}
Note that there is no term linear in $\mu\equiv 2|\alpha|^2$ -- that is, if we ignore the multi-photon contribution, the two states are indistinguishable. This happens because, to linear order in $\alpha$, the two-dimensional subspace spanned by the two $x$-basis signals and the two-dimensional subspace spanned by the $y$-basis signals coincide. The large overlap of $|\Psi_x\rangle$ and $|\Psi_y\rangle$ indicates that, averaged over the value of the key bit, the states emitted by the source when the $x$ basis is chosen are hard to distinguish from the states emitted when the $y$ basis is chosen. (We have chosen the purifications $|\Psi_x\rangle$ and $|\Psi_y\rangle$ that have the maximal overlap compatible with the key-averaged signal states \cite{norbert}; that is why we have paired $|1_Y\rangle$ with $|i\alpha\rangle$ and  $|0_Y\rangle$ with $ |-i\alpha\rangle$ in eq.~(\ref{Psi-y}).)

\begin{figure}
\begin{center}
\setlength{\unitlength}{1.1pt}
\begin{picture}(350,73)
\put(50,0){\framebox(20,20){\scriptsize{$~$Eve} }}
\put(90,0){\framebox(20,20){{\scriptsize Filter}}}
\put(10,20){\framebox(20,20){$~{|\Psi_{x}\rangle}$ }}

\put(144,4){\makebox(0,12){\shortstack{{\scriptsize Bob}\\{\scriptsize measures $M_{x}$}}}}
\put(74,44){\makebox(0,12){\shortstack{{\scriptsize Alice}\\{\scriptsize measures $x$}}}}

\put(24,50){\makebox(0,12){\scriptsize qubit}}
\put(24,-2){\makebox(0,12){\scriptsize mode}}

\put(30,10){\line(1,0){20}}
\put(70,10){\line(1,0){20}}
\put(110,10){\line(1,0){10}}
\put(30,10){\line(-1,1){10}}
\put(20,40){\line(1,1){10}}
\put(30,50){\line(1,0){20}}
%
\put(100,26){\line(1,0){6}}
\put(100,20){\line(0,1){6}}

\put(230,0){\framebox(20,20){{\scriptsize $~$Eve }}}
\put(270,0){\framebox(20,20){{\scriptsize Filter}}}
\put(190,20){\framebox(20,20){$~{|\Psi_{y}\rangle}$ }}

\put(324,4){\makebox(0,12){\shortstack{{\scriptsize Bob}\\{\scriptsize measures $M_{y}$}}}}
\put(254,44){\makebox(0,12){\shortstack{{\scriptsize Alice}\\{\scriptsize measures $y$}}}}

\put(204,50){\makebox(0,12){\scriptsize qubit}}
\put(204,-2){\makebox(0,12){\scriptsize mode}}

\put(210,10){\line(1,0){20}}
\put(250,10){\line(1,0){20}}
\put(290,10){\line(1,0){10}}
\put(210,10){\line(-1,1){10}}
\put(200,40){\line(1,1){10}}
\put(210,50){\line(1,0){20}}
%
\put(280,26){\line(1,0){6}}
\put(280,20){\line(0,1){6}}
\end{picture}

\vskip .1in
($a$) \hskip 2.55truein
($b$)
\vspace*{0.1truein}
\end{center}
\fcaption{Roles of Alice, Bob, and Eve in the key distribution protocol. Alice determines her key bit by performing an ideal measurement on a qubit that is entangled with the signal  mode. ($a$) If Alice declares the $x$ basis, then Alice measures in the $x$ basis and the entangled state of her qubit and the mode is $|\Psi_x\rangle$. ($b$) If she declares the $y$ basis, she measures in the $y$ basis and the entangled state is $|\Psi_y\rangle$. Bob's detector applies a basis-independent filter that determines whether his measurement has a conclusive outcome, followed by a two-outcome POVM, either $M_x$ or $M_y$ depending on Bob's declared basis. If Alice and Bob both declare the $x$ basis, the probability that their measurement outcomes disagree is $\delta_x$, and if both declare the $y$ basis, the probability that they disagree is $\delta_y$.}
\label{fig:setup}
\end{figure}

A protocol such that the signals emitted by the source, averaged over the key bit, have a small dependence on Alice's basis choice can be analyzed by considering an equivalent protocol in which the basis-dependence of the signals can be related to the ``balance'' of a ``quantum coin'' \cite{GLLP,koashi}. In this equivalent protocol, Alice measures the coin in the basis $\{|0_Z\rangle, |1_Z\rangle\}$ to determine whether her signal is encoded in the $x$ basis or the $y$ basis, and we may take the joint state of the coin and Alice's source states to be 
\begin{equation}
|\Phi\rangle= \left(|0_Z\rangle\otimes |\Psi_x\rangle + |1_Z\rangle\otimes|\Psi_y\rangle \right)/\sqrt{2}~;
\end{equation}
furthermore we may imagine that the measurement of the coin is delayed until after Eve is finished interacting with the signals. 

If $|\Psi_x\rangle$ and $|\Psi_y\rangle$ were equal, then the signals, averaged over the two possible values of the key bit, would be independent of whether Alice chose the $x$ basis or the $y$ basis. If $|\Psi_x\rangle$ and $|\Psi_y\rangle$ are nearly equal, then the source leaks a small amount of information about Alice's basis choice that Eve might exploit. A useful way to quantify the leaked basis information is to consider what would happen if Alice were to measure each of her coins in the basis of $X$ eigenstates rather than the basis of $Z$ eigenstates; then the outcome $X=-1$ would occur with probability $\Delta$, where
\begin{equation}
\langle \Psi_y|\Psi_x\rangle = 1-2\Delta~.
\end{equation}
Thus we say that the coin is ``$\Delta$-balanced,'' where $\Delta$ quantifies the basis dependence of Alice's signals. For the signal states eq.~(\ref{bb84states}), we have
\begin{eqnarray}
\label{Delta}
\Delta&=&\frac{1}{2}\left(1- e^{-\mu/2}\left(\cos(\mu/2) + \sin (\mu/2)\right)\right)\nonumber\\
&=& \mu^2/8 +O(\mu^3)~.
\end{eqnarray}

However, not all of the signals emitted by the source are detected; Eve, by carefully choosing which signals to block, might be able to enhance the basis dependence of the detected signals. Let us pessimistically assume that the detected signals are chosen to maximize the imbalance of the coin. For a perfect channel and for perfect detection efficiency, the fraction of signals detected would be $1-e^{-\mu}\approx \mu$, where $\mu$ is the mean photon number of the signals emitted by the source. The actual fraction of detected signals is $\eta \mu $, where $\eta$ is an effective rate of loss due to absorption in the channel and imperfect detector performance. In the worst case, for all of the signals that are removed, the coin is an $X=1$ eigenstate, which would enhance the probability of the outcome $X=-1$ by
\begin{equation}
\label{delta-with-loss}
\Delta \to \Delta' =  \Delta/(\eta\mu)\approx \mu/(8\eta)~.
\end{equation}
Thus we will use $\Delta'$ to quantify the basis dependence in the signals that determine Alice's sifted key.

We will argue that for a fixed $\Delta'$ that is sufficiently small, secure key can be extracted from sifted key at a fixed nonzero asymptotic rate. Therefore, the number of private key bits that can be generated per transmitted signal is proportional to $\eta\mu\approx 8\Delta'\eta^2$ and hence scales like $\eta^2$ --- the same scaling as for a source with random phases \cite{ilm,GLLP} (assuming that decoy states are not used). 

\subsection{The uncertainty principle and privacy amplification}
Koashi \cite{koashi}, generalizing the techniques in \cite{GLLP}, explained how to relate the imbalance of the quantum coin to the privacy of a key that is generated by performing ideal measurements on qubits; here we wish to apply his reasoning to Alice's key bits, which we regard as determined by outcomes of measurements of (fictitious) qubits entangled with the signal states. For our argument, it is important that the imbalance of the coin arises solely from the basis dependence of Alice's signals. That is why we required that Bob's measurement be modeled by a basis-independent filter followed by a two-outcome measurement; if the probability of an inconclusive outcome actually depended on the basis, then the basis-dependence of the filter might have further enhanced the imbalance of the coins. As it is, after the inconclusive measurement results are discarded, we may imagine that Bob performs his two-outcome POVM on signals that are entangled with Alice's qubits and the quantum coins, where the imbalance $\Delta'$ of the coins is determined only by the basis dependence of the source and the loss in the channel.

We recall that the BB84 protocol includes a verification test conducted by Alice and Bob. In the test, Alice and Bob publicly compare a randomly selected subset of their key bits to estimate a bit error rate, the fraction of the key bits for which Alice and Bob disagree. Signals sent in the $x$ basis and the $y$ basis can be tested separately, to estimate both a bit error rate $\delta_x$ for signals sent in the $x$ basis, and a bit error rate $\delta_y$ for signals sent in the $y$ basis. Let us consider the key bits that are generated by measuring in the $x$ basis. Using Koashi's method \cite{koashi}, we may show that, in the asymptotic limit of a very long key, 
private key can be extracted from Alice's sifted key at a rate 
\begin{equation}
\label{key-rate}
R=1- H(\delta_x) - H(\delta_y')~,
\end{equation}
where $\delta_x$ is the bit error rate, $\delta_y'\ge \delta_y$ is a function of $\delta_y$ and $\Delta'$, and  $H(\delta)= -\delta\log_2\delta - (1-\delta)\log_2(1-\delta)$ is the binary Shannon entropy. A similar formula applies for the key bits generated by measuring in the $y$ basis, but with $\delta_x$ and $\delta_y$ interchanged. The term $H(\delta_x)$ is the rate at which sifted key must be sacrificed to perform error correction that reconciles Bob's key with Alice's (according to standard Shannon theory), and $H(\delta_y')$ is the cost of performing privacy amplification to ensure that Eve has negligible information about Alice's final key.

\begin{figure}
\begin{center}
\setlength{\unitlength}{1.4pt}
\begin{picture}(170,73)
\put(50,0){\framebox(20,20){$~$Eve }}
\put(90,0){\framebox(20,20){Filter}}
\put(10,20){\framebox(20,20){$~{|\Psi_{x}\rangle}$ }}

\put(144,4){\makebox(0,12){\shortstack{{\scriptsize Bob}\\{\scriptsize measures $M_{y}$}}}}
\put(74,44){\makebox(0,12){\shortstack{{\scriptsize Alice}\\{\scriptsize measures $y$}}}}

\put(24,50){\makebox(0,12){qubit}}
\put(24,-2){\makebox(0,12){mode}}

\put(30,10){\line(1,0){20}}
\put(70,10){\line(1,0){20}}
\put(110,10){\line(1,0){10}}
\put(30,10){\line(-1,1){10}}
\put(20,40){\line(1,1){10}}
\put(30,50){\line(1,0){20}}
%
\put(100,26){\line(1,0){6}}
\put(100,20){\line(0,1){6}}

\end{picture}
\vspace*{0.1truein}
\end{center}
\fcaption{Setting that defines the ``phase error rate'' $\delta_y'$. The entangled state of Alice's qubit and the signal mode is $|\Psi_x\rangle$ as though Alice had declared the $x$ basis, but Alice and Bob both measure in the ``wrong'' basis --- Alice measures $y$ instead of $x$, and Bob measures $M_y$ instead of $M_x$. The probability that Bob's outcome disagrees with Alice's is $\delta_y'$. The quantity $\delta_y'$ cannot be directly measured in the protocol, but an upper bound can be inferred from the observed bit error rate $\delta_y$ and the known basis dependence of Alice's signal states. }
\label{fig:delta-y-prime}
\end{figure}

What is the quantity $\delta_y'$? Recall that when the $x$ basis is declared, Alice measures her qubit in the $x$ basis and Bob performs measurement $M_x$, while if the $y$ basis is declared, Alice measures her qubit in the $y$ basis and Bob performs measurement $M_y$. Furthermore, when the $x$ basis is declared, the entangled state of Alice's qubit and the signal mode is $|\Psi_x\rangle$, while when the $y$ basis is declared, the entangled state is $|\Psi_y\rangle$. The quantity $\delta_y'$ is an upper bound on what the bit error rate would have been if Alice measured in the $y$ basis and Bob performed measurement $M_y$, but where the entangled state of the qubit and mode is $|\Psi_x\rangle$ instead of $|\Psi_y\rangle$. This hypothetical error rate is not known directly from the test, but because the state $|\Psi_x\rangle$ is close to $|\Psi_y\rangle$ when $\Delta$ is small, the hypothetical ``phase error rate'' $\delta_y'$ is close to the actual observed bit error rate $\delta_y$. See Fig.~\ref{fig:delta-y-prime}.

The estimate of $\delta_y'$ is explained in Appendix \ref{app:uncertainty}. The result is that for any $\varepsilon>0$, with high probability we may express $\delta_y'$ as
\begin{eqnarray}
\label{delta-y-prime}
\delta_y' ~= &&\delta_y + 4\Delta'(1-\Delta')(1-2\delta_y) + ~4(1-2\Delta')\sqrt{\Delta'(1-\Delta')\delta_y(1-\delta_y)}+\varepsilon\nonumber\\
&& \le ~\delta_y + 4\Delta' +4\sqrt{\Delta'\delta_y} +\varepsilon~.
\end{eqnarray} 
The upper bound on $\delta_y'$ in the last line is obtained by retaining only the terms of lowest order in $\delta_y$ and $\Delta'$, and gives a reasonable approximation to the tighter bound on the previous line. From the tighter bound and eq.~(\ref{key-rate}), we see that the key generation rate $R$ remains positive, in the limit of negligible bit error rates $\delta_x$, $\delta_y$, for $\Delta' < 0.146$, which corresponds to $\mu$ smaller than about $(1.16) \eta$ when $\eta$ is small.

Why does $\delta_y'$ characterize the cost of amplifying privacy? Following Koashi \cite{koashi} (and generalizing an idea used in \cite{mayers} and \cite{koashi-preskill}) we observe that when the $x$ basis is declared, Eve's knowledge about Alice's key would not have been affected if Bob had chosen to measure in the wrong basis (to measure $M_y$ rather than $M_x$). Furthermore, neither the key nor Eve's knowledge of the key would have been affected if Alice had chosen to delay measuring her qubits until after Bob measured the signals that he received. Therefore, we may imagine that Alice generated her key by measuring in the $x$ basis a string of qubits that, conditioned on the outcome of Bob's $M_y$ measurement, are close to $y$-basis eigenstates (if $\delta_y'$ is small). 

If, after Bob's measurement, each of Alice's qubits were {\em exactly} a $y$-basis eigenstate, then Eve would be powerless to predict the key bits that Alice obtains by measuring the qubits in the $x$ basis. In that case we could say that the privacy of the key is founded on the uncertainty principle: a qubit that produces a deterministic outcome when measured in the $y$ basis will produce a uniformly random outcome when measured in the $x$ basis. From this perspective, we may say that the purpose of the privacy amplification is to ensure that Alice's final key is very nearly equivalent to a key determined by measuring $y$-basis eigenstates in the $x$ basis.

Let the $n$-component binary vector $v$ denote Alice's sifted key for the signals in which the $x$ basis was declared by Alice and Bob, and Bob reported successfully detecting the signal. Alice extracts her $k$-bit final key $\kappa= vG$ by applying the random rank-$k$ binary $n\times k$ matrix $G$ to the sifted key. Equivalently, we may say that Alice's final key is obtained by measuring the $k$ binary observables
\begin{equation}
\bar X_i=\bigotimes_{j: G_{ji}=1} X_j~, \quad i=1,2,3,\dots k~,
\end{equation}
where $X_j$ denotes the Pauli matrix $X$ acting on Alice's $jth$ qubit.

Before selecting the matrix $G$, we might have first selected a random rank-$(n-k)$ binary $n\times(n-k)$ matrix $H$, and then chosen $G$ subject to the constraint $G^TH=0$ (so that each column of $G$ is orthogonal in the binary inner product to each column of $H$). In that case, each of the $n-k$ binary observables
\begin{equation}
\bar Y_m=\bigotimes_{j: H_{jm}=1} Y_j~, \quad m=1,2,3,\dots n-k~
\end{equation}
(where $Y_j$ denotes the Pauli matrix $Y$ acting on Alice's $jth$ qubit) would commute with each of the $\bar X_i$'s. We may therefore imagine (since it would have no effect on Alice's final key or on Eve's information about the key) that Alice measured each of the $\bar Y_m$'s before measuring the $\bar X_i$'s to determine her final key bits.

Now, if Alice were to measure all of her qubits in the $y$ basis, then with probability exponentially close to 1 for $n$ large, her outcome would be one of $N_{\rm typ}=2^{n(H(\delta_y')+\varepsilon)}$ typical $n$-bit strings. Note that if $u$ and $v$ are two distinct $n$-bit strings, and $s$ is a randomly chosen $n$-bit string, then the bits $us^T$and $vs^T$ are distinct with probability 1/2. Therefore if we choose the number of randomly chosen binary observables in the set $\{\bar Y_m\}$ (the number of columns in the matrix $H$) to be
\begin{equation}
n-k = n\big(H(\delta_y')+2\varepsilon\big)~,
\end{equation}
then the probability that more than one typical binary string is compatible with the outcomes of the $\{\bar Y_m\}$ measurements is at most $N_{\rm typ}\cdot 2^{k-n}= 2^{-\varepsilon n}$. This means that, conditioned on the results of measuring all the observables in the set $\{\bar Y_m\}$, Alice's $n$ qubits have been projected to a state that is exponentially close to a product of $y$-basis eigenstates. Thus Alice's final key is essentially the same as it would have been if the sifted key had been generated by measuring $y$-basis eigenstates in the $x$ basis, and therefore is guaranteed to be private. Since $k= n\big(1-H(\delta_y')-2\varepsilon\big)$, and $n\big(H(\delta_x)+\varepsilon'\big)$ key bits are sacrificed to reconcile Bob's sifted key with Alice's, we obtain the asymptotic key rate eq.~(\ref{key-rate}). 

\subsection{Universal composability}
This argument shows not just that Eve's mutual information with the final key is negligible, but also that a stronger definition of security is satisfied: the protocol is {\em universally composable} \cite{ben-or, renner-konig,konig-renner}. This means that the final key bits can be safely used in any subsequent application without compromising security. To demonstrate composability, we consider in more depth (following \cite{konig-renner}) the joint state shared by Alice and Eve before Alice performs her $x$-basis measurements to generate her key. We have already seen that, after Alice's fictitious measurement of the observables $\{\bar Y_m\}$, her density operator $\rho_A$ has high fidelity with the pure state $|y_0\rangle$, where $|y_0\rangle$ denotes a particular product of $y$-basis eigenstates:
\begin{equation}
F(\rho_A, |y_0\rangle\langle y_0|)\equiv \langle y_0|\rho_A|y_0\rangle = 1- \zeta^2~,
\end{equation}
where $\zeta$ is exponentially small. Therefore we may introduce an auxiliary system $E$, which we pessimistically assume to be under the eavesdropper's control, such that $\rho_A$ has a purification $|\Upsilon\rangle_{AE}$ where $|\Upsilon\rangle_{AE}$ has a large overlap with $|y_0\rangle_A\otimes |\lambda\rangle_E$ and $|\lambda\rangle_E$ is a pure state of $E$: 
\begin{equation}
F\big(\left(|\Upsilon\rangle\langle\Upsilon|\right)_{AE}, \left(|y_0\rangle\langle y_0|\right)_{A}\otimes\left(|\lambda\rangle \langle \lambda| \right)_{E}\big)= 1-\zeta^2~.
\end{equation}
We can relate the large overlap to proximity in the trace norm, using the inequality
\begin{equation}
\frac{1}{2}\| \rho - \sigma\|_{\rm tr}\le \sqrt{1-F(\rho,\sigma)}~,
\end{equation}
obtaining
\begin{equation}
\frac{1}{2}\|  \left(|\Upsilon\rangle\langle \Upsilon| \right)_{AE} - ~\left(|y_0\rangle\langle y_0|\right)_A\otimes \left(|\lambda\rangle\langle \lambda|\right)_E \|_{\rm tr} \le \zeta~.
\end{equation}

Now when Alice measures the observables $\{\bar X_i\}$ to determine her final key, the state $|y_0\rangle\langle y_0|$ yields a uniformly random key described by the maximally mixed density operator 
\begin{equation}
\rho_A^{\rm uni}=\frac{1}{2^k}\sum_{\kappa=0}^{2^k-1} |\kappa\rangle\langle \kappa |~,
\end{equation}
and the resulting state shared by Alice and Eve is a product state $\rho_A^{\rm uni}\otimes \sigma_E$, where $\sigma_E=\left(|\lambda\rangle\langle \lambda|\right)_E$. Quantum operations cannot increase the trace distance; therefore, if $\rho_{AE}$ is the state shared by Alice and Eve after Alice's measurement, we find that
\begin{equation}
\label{composable} 
\frac{1}{2}\| \rho_{AE} - \rho_A^{\rm uni}\otimes \sigma_E\|_{\rm tr} \le \zeta~.
\end{equation}
Eq.~(\ref{composable}) is a security criterion shown to be universally composable in \cite{renner-konig}, which we have now seen is satisfied by a protocol where the signal states have nonrandom phases. This demonstration of  universal composability for the BB84 quantum key distribution protocol also applies to the other cases addressed by Koashi in \cite{koashi,koashi-realistic}.

\section{Bright reference pulse}
\label{sec:bright}
Our security proof also applies if the reference pulse is bright. Suppose that, for each of the BB84 signals, the state of the reference pulse emitted by the source is $|\beta\rangle_R$, while the emitted state of the signal pulse is as in eq.~(\ref{bb84states}), where $|\beta| \ge |\alpha|$. Then we may calculate the imbalance of the quantum coin as before, except in eq.~(\ref{Delta}) we replace $\mu/2$ by $\mu_S=|\alpha|^2$, the mean photon number of the signal pulse. When Bob receives the signals, he first attenuates the reference pulse so that its strength matches the strength of the signal pulse, and then performs the interferometric measurement described earlier. Thus we replace $\eta\mu$ in eq.~(\ref{delta-with-loss}) by $2\eta\mu_S$, the mean total photon number of the two modes after the attenuation of the reference pulse (and taking loss into account). Therefore, in both formulas we replace $\mu$ by $2\mu_S$, finding
\begin{equation}
\Delta'\approx \mu_S/(4\eta)~,
\end{equation}
when the signal pulse is weak. It is obviously essential for this argument that the (possibly bright) reference pulse has exactly the same state (or at least very nearly the same state) for each of the BB84 signals, so that it conveys no information (or very little information) to Eve concerning which signal is emitted. Eve might tamper with Bob's measurement of the signal by attacking the reference pulse, but because she knows little about the basis in which the key bit is encoded, she is still incapable of collecting much information about the key without creating a detectable disturbance.

Why is it that Eve seems to be no better off when the reference pulse is bright than when it is as dim as the signal pulse? The reason is that even when the reference pulse is dim we have pessimistically assumed that Eve knows the phase of the reference pulse perfectly. When we say that Eve ``knows'' that the state of the reference pulse is the coherent state $|\beta\rangle_R$, we are taking it for granted that the phase of $\beta$ has a fixed value relative to a classical phase reference that Eve controls. If the phase of $\beta$ were not already known, then increasing $|\beta|$ would make it easier for Eve to determine the phase relative to her reference frame, but if the phase is already known then increasing $|\beta|$ does not give Eve any additional advantage.

The observation that the reference pulse can be bright without compromising security is relevant to, for example, the ``double Mach-Zehnder'' scheme for implementing phase encoding \cite{b92,gisin}. In this set-up, Alice uses an unbalanced interferometer to split a single pulse into two spatially separated pulses that travel to Bob through the same optical fiber; one of these pulses, whose phase is unmodulated, is the reference pulse, and the other, whose phase is modulated, is the signal pulse. Bob uses another unbalanced interferometer to combine the two pulses, with a modulated relative phase that determines his measurement basis. 

For a given strength of the signal pulse, using a brighter reference pulse increases the key rate by reducing the probability that Bob fails to detect the signal. In particular, if the reference pulse is much brighter than the signal pulse, then nearly twice as many signals will be detected compared to the case where the reference and signal pulses have equal strength. Our proof shows that the key rate can be doubled by using a bright reference pulse, without reducing the privacy of the final key. 

We can understand this doubling of the key rate by considering how Bob's detection system operates in the double Mach-Zehnder scheme. Let us suppose that the reference pulse follows the longer path in Alice's unbalanced interferometer and therefore lags behind the signal pulse during transmission through the fiber; we therefore use $a_L$ ($L$ for ``long'') to denote the reference mode, and $a_S$ ($S$ for ``short'' as well as ``signal'') to denote the signal mode. The two paths in Bob's interferometer also have unequal lengths, to compensate for the spatial separation in the fiber of the reference and signal pulses: the portion of the reference pulse that follows the short path in Bob's interferometer (let us call it the $LS$ mode, since it follows the long path in Alice's interferometer and the short path in Bob's) interferes with the portion of the signal pulse that follows the long path in Bob's interferometer (the $SL$ mode). The first beam splitter in Bob's interferometer is asymmetric, so that the reference and signal pulses are split according to
\begin{eqnarray}
&& |\tilde \beta\rangle_L \mapsto |c\tilde \beta\rangle_{LL}\otimes |s\tilde\beta\rangle_{LS}~,\nonumber\\
&& |\tilde\alpha\rangle_S \mapsto |c\tilde\alpha\rangle_{SL}\otimes |s\tilde \alpha\rangle_{SS}~,
\end{eqnarray}
where $c^2 + s^2=1$. Here $LL$ refers to the part of the reference pulse that enters the long arm of Bob's interferometer, and $SS$ refers to the part of the signal pulse that enters the short arm; these $LL$ and $SS$ modes are ignored by Bob. We denote the strengths of the reference and signal pulses by $\tilde\beta$ and $\tilde\alpha$ rather than $\beta$ and $\alpha$ to take into account the attenuation of the pulses during transmission through the fiber; thus, if Eve does not intervene, $|\tilde\alpha|/|\tilde\beta|=|\alpha|/|\beta|$.

Bob's first beam splitter is chosen so that $s |\tilde\beta|= c|\tilde \alpha|$, or $s/c=|\alpha|/|\beta|$; therefore the incoming $LS$ and $SL$ modes at Bob's second beam splitter have equal strength. This second beam splitter is symmetric, and Bob reads out his device by detecting the photons in the two output ports of this symmetric beam splitter. The probability of detection is proportional to the total mean photon number in the $LS$ and $SL$ modes as they arrive at Bob's second beam splitter, which is 
\begin{equation}
s^2|\tilde\beta|^2 + c^2 |\tilde\alpha|^2= 2c^2|\tilde\alpha|^2~,
\end{equation}
where
\begin{equation}
 c^2 = \frac{1}{1+s^2/c^2}= \frac{|\tilde\beta|^2}{|\tilde\alpha|^2 +|\tilde\beta|^2}=\frac{|\beta|^2}{|\alpha|^2 +|\beta|^2}~.
\end{equation}
The  factor $c^2$ suppressing the detection rate arises because the strength of the signal pulse following the long path in Bob's interferometer is attenuated by the factor $c$ at the first beam splitter; the rest of the signal pulse follows the short path and is wasted. This factor $c^2$ is $1/2$ for $|\alpha|=|\beta|$ but close to 1 for $|\beta|^2 \gg |\alpha|^2$. Thus when the reference pulse and signal pulse have the same strength half of the signal is wasted, but when the reference pulse is bright hardly any of the signal is wasted.

\section{Comparison to phase-randomized signals}
\label{sec:comparison}
It is instructive to compare our estimate of the key generation rate with the rate that would be achievable if the phases of the signals were random. In that case, the quantum coin is undisturbed if the source emits a single photon, but it is strongly affected if two or more photons are emitted. Thus the fraction of the coins that are damaged is at worse $\Delta'=p_M/p_D$, where $p_M$ is the probability for emission of a multi-photon, and $p_D$ is the fraction of the signals that are detected. Therefore, for small $\mu$ we have \cite{GLLP}
\begin{equation}
\Delta'= p_M/p_D\approx \frac{{1\over 2} \mu^2}{\eta \mu}= \mu/2\eta~.
\end{equation}
The key generation rate again has the form eq.~(\ref{key-rate}), but for coins in this particular type of state (a fraction $\Delta'$ badly damaged and a fraction $1-\Delta'$ undisturbed), the estimate of $\delta_y'$ can be improved to
\begin{equation}
\label{delta-p-tagged}
\delta_y' -\delta_y  =\Delta'/2 \approx \mu/4\eta~.
\end{equation} 
(Eq.~(\ref{delta-p-tagged}) is an improvement found in \cite{boileau} of the result $\delta_y'-\delta_y =\Delta'$ found in \cite{GLLP}.) For comparison, by combining eq.~(\ref{delta-with-loss}) with eq.~(\ref{delta-y-prime}) we find
\begin{equation}
\delta_y' -\delta_y\approx \mu/2\eta + \sqrt{2\mu\delta_y/\eta}
\end{equation} 
for the case where Eve knows the phase of the signals; thus we see that phase randomization improves the rate, but only by a relatively modest amount when the bit error rate is small.

In fact, as shown in \cite{GLLP}, for a phase-randomized source a higher key generation rate can be achieved than implied by eq.~(\ref{key-rate}) and eq.~(\ref{delta-p-tagged}), furthermore as shown in \cite{koashi-realistic} this higher rate applies for the detector model assumed in Sec.~\ref{sec:modeling}. The achievable number $G$ of key bits {\em per pulse} can be expressed as
\begin{equation}
\label{random-key-per-pulse}
G = \frac{1}{2}\Big( Q_1\big(1-H(e_1)\big) -Qf(e)H(e)\Big)~.
\end{equation}
Here $Q$ is the fraction of pulses for which Bob detects the signal, $e$ is the bit error rate observed in the verification test, and $f(e)\ge 1$ parametrizes the inefficiency of the error correction used to perform key reconciliation; furthermore $e_1$ is the error rate for the single-photon signals emitted by the source and $Q_1$ is the fraction of the single-photon signals that result in a detection event. (The factor of $\frac{1}{2}$ in front of the expression for $G$ arises because half of the detected signals, those for which Alice and Bob used different bases, are discarded during sifting.) In a protocol that uses decoy states, $Q_1$ and $e_1$ can be estimated directly, but if decoy states are not used then we may pessimistically assume all the bit errors are due to single-photon signals, so that 
\begin{eqnarray}
e_1 = e/(1-\Delta')~
\end{eqnarray}
(with $\Delta'=p_M/Q$), and that all the multi-photon signals are detected, so that 
\begin{eqnarray}
Q_1=Q(1-\Delta')~.
\end{eqnarray} 

\begin{figure}
\begin{center}
\epsfig{file=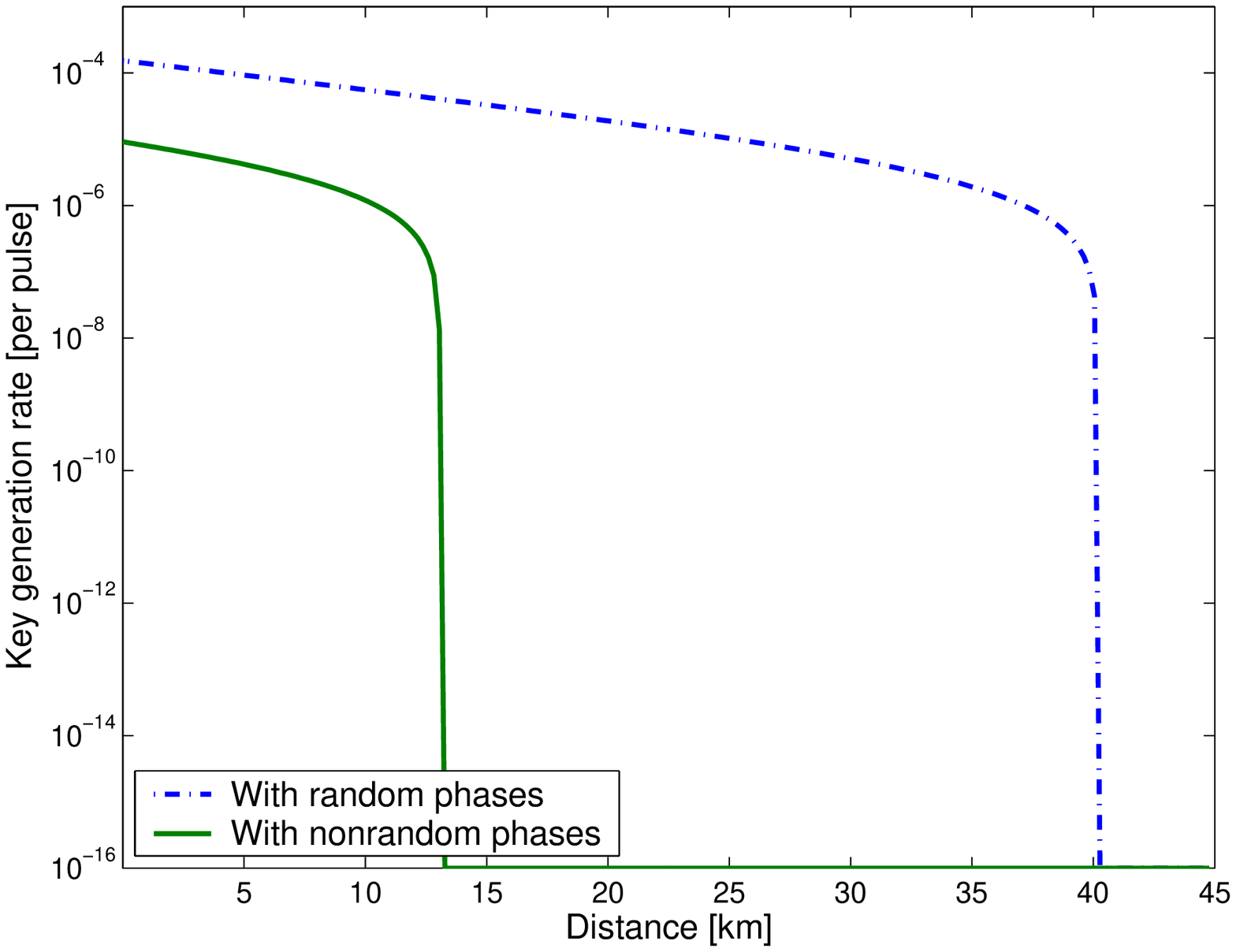,width=10cm} 
\end{center}
\fcaption{The key generation rate in bits per pulse as a function of distance for sources with random and nonrandom phases, using experimental parameters from \cite{GYS}, and assuming an error-correction inefficiency $f(e)=1.22$. Here for the phase-randomized case we assume that decoy states are not used. For each value of the distance, the signal strength has been chosen to optimize the rate. } 
\label{fig:plot}
\end{figure}

For the case of a source with nonrandom phases, the achievable key rate per pulse is
\begin{equation}
\label{nonrandom-key-per-pulse}
G = \frac{1}{2} Q\big(1-f(e)H(e) - H(e_{\rm ph})\big) ~;
\end{equation}
here $e_{\rm ph}$ as a function of $e$ is given as in eq.~(\ref{delta-y-prime}) by
\begin{eqnarray}
e_{\rm ph} = e + 4\Delta'(1-\Delta')(1-2e)
+ ~4(1-2\Delta')\sqrt{\Delta'(1-\Delta')e(1-e)}~,
\end{eqnarray} 
where 
\begin{eqnarray}
\Delta' = \Delta/Q =\frac{1}{2Q}\left(1- e^{-\mu/2}\left(\cos(\mu/2) + \sin (\mu/2)\right)\right)~.
\end{eqnarray}
We have plotted the key rates per pulse given by eq.~(\ref{random-key-per-pulse}) and eq.~(\ref{nonrandom-key-per-pulse}) as a function of distance in Fig.~\ref{fig:plot}. For this comparison, we have used experimental parameters (channel loss, dark count rate, and detector efficiency) from \cite{GYS}, and assumed an error-correction inefficiency $f(e)=1.22$; for each value of the distance, the signal strength $\mu$ has been chosen to optimize the rate. The maximal distance for which secure key exchange is possible is about three times longer in the phase-randomized case.

\section{Phase modulated reference pulse}
\label{sec:ukd}

The security proof works because the overlap of $|\Psi_x\rangle$ and $|\Psi_y\rangle$ is $1-O(\mu_S^2)$. This property holds because, if we assume that the phase of the reference pulse is unmodulated, then the four BB84 signal states span a two-dimensional space when the multi-photons in the signal pulse are neglected. But, even when the reference pulse is weak,  the story can change if the source modulates the phase of the reference pulse as well as the signal pulse. In \cite{lo-jp-attack} we studied such a source, and we concluded that the eavesdropper can exploit the modulation of the reference pulse to improve the effectiveness of her attack. 

In particular, in Appendix \ref{app:ukd} (see also \cite{lo-jp-attack}), we study the BB84 signal states 
\begin{eqnarray} \label{bb84states-lojp}
\ket{\tilde 0_X} &=& \ket{\alpha}_R\otimes \ket{\alpha}_S~, \nonumber \\
\ket{\tilde 1_X} &=& \ket{-i\alpha}_R\otimes \ket{i\alpha}_S~, \nonumber \\
\ket{\tilde 0_Y} &=& \ket{e^{-i\pi/4}\alpha}_R\otimes \ket{e^{i\pi/4}\alpha}_S~, \nonumber \\
\ket{\tilde 1_Y} &=& \ket{e^{i\pi/4}\alpha}_R\otimes \ket{e^{-i\pi/4}\alpha}_S~, 
\end{eqnarray}
in which the reference pulse for each signal has a distinct phase relative to Eve's classical phase reference. Thus key information is encoded not just in the relative phase of the signal and reference pulses, but also in the phase of the (dim) reference pulse relative to Eve's reference frame. The signal set eq.~(\ref{bb84states-lojp}) seems more natural if we recognize that, when re-expressed in a different basis, it becomes the signal set eq.~(\ref{bb84states-old}); thus it arises in a polarization encoding scheme where the single-photon component of each signal is the standard BB84 single-qubit signal, and where the relative phase of the vacuum and single-photon component is the same for all four signals. 

We note that, even if we ignore the (small) multi-photon component of the weak coherent state signals, these four BB84 signals belong to a three-dimensional Hilbert space spanned by the vacuum (no-photon) state and two distinguishable one-photon states. On the other hand, the two signals that convey a particular key bit value (0 or 1) span a two-dimensional space. Therefore, an eavesdropper who knows the relative phase of the vacuum and single photon component can launch an attack on the protocol that we call {\em unambiguous key discrimination}. She performs a POVM with three outcomes: 0, 1, and inconclusive; if either of the conclusive outcomes occurs, then Eve knows with certainty the key bit (0 or 1) encoded in the BB84 signal emitted by the source, though she does not gain any information about which of the two possible BB84 states compatible with that key bit was emitted. 

For these signal states, we show in Appendix \ref{app:ukd} that Eve's key discrimination has a conclusive outcome with probability $p_C\approx .146\mu$ when the mean photon number $\mu$ is small. If the signals had random phases, then only the multi-photon signals would be insecure, a fraction  $O(\mu^2)$ of all the signals. But when the phase of $\alpha$ is known and the signals eq.~(\ref{bb84states-lojp}) are used, this fraction increases to $O(\mu)$.

As van Enk \cite{vanEnk} has astutely pointed out to us, the unambiguous key discrimination attack becomes much less effective if the encoding of the signals is chosen judiciously. And for the signal set eq.~(\ref{bb84states}) in which the reference pulse has the same phase in all four signals, the signals span a two-dimensional space when the multi-photons are neglected, so that the conclusive key discrimination succeeds with probability $p_C=O(\mu^2)$. (For polarization encoding, this corresponds to particular choices for the phases of the coherent states and for the orientation of the BB84 polarization states in the Bloch sphere.) Thus as we have seen, unconditional security is provable for this particular encoding, but the proof does not apply to the signal set eq.~(\ref{bb84states-lojp}). Therefore, the protocol is vulnerable to a ``Trojan horse'' attack, in which a malicious manufacturer in cahoots with Eve provides Alice with a source that modulates the phase of the reference pulse \cite{norbert-private}. 

There are other possible Trojan horse attacks that could be even more devastating --- for example the source might encode the four BB84 states in four distinct bosonic modes so that Eve can distinguish the signals perfectly. It is useful, however to differentiate such multi-mode Trojan horse attacks from the more subtle attacks where the protocol is compromised because the source modulates the phase of a single mode. The single-mode Trojan horse might be harder for Alice and Bob to detect, and might be less susceptible to countermeasures \cite{trojan} such as filtering out the unwanted modes. 

\section{Conclusion}
\label{sec:conclusion}
In summary, we have shown that the BB84 quantum key distribution protocol with phase-encoded signals is secure if the signal pulse is weak. Our proof applies even if the reference pulse is bright and has a phase known by the adversary, provided that the phase of the reference pulse is the same for all four of the BB84 signals, and the measurement device satisfies the basis-independent filter assumption formulated in Sec.~\ref{sec:modeling}. Furthermore, our proof shows that the BB84 protocol is universally composable under these conditions. Our security proof also applies to polarization encoding, if the signal states are chosen appropriately. And we have discussed a new type of {\em single-mode} Trojan horse attack, in which security can be compromised because the source modulates the phase of the reference pulse depending on which signal is being emitted. 

The achievable key rate established by our proof scales quadratically with the transmission $\eta$ in the channel, as for BB84 with phase-randomized weak coherent-state signals. In the phase-randomized case, a key rate linear in $\eta$ can be achieved with the decoy state method. An interesting open question is whether employing decoy states can also extend the range over which secure BB84 quantum key distribution is possible in the case where the eavesdropper knows the phase of the (possibly bright) reference pulse and the detector is modeled as in Sec.~\ref{sec:modeling}. We note that a key rate linear in $\eta$ has been established for the B92 \cite{b92} protocol assuming that the reference pulse is bright and the detector has suitable properties (one proof requires Bob to have a local oscillator phase-locked with the reference pulse \cite{koashi-local-oscillator}, another requires that Bob's detector can distinguish single-photon signals from multi-photons \cite{b92-bright}), but these proofs do not apply to the threshold detectors used in current experiments. It may also be fruitful to investigate the impact on security of using imperfect sources and detectors for other quantum key distribution protocols, such as the six-state protocol \cite{six-state} and SARG04 \cite{sarg}.

Finally we remark that, since a higher key generation rate has been established for the case where the reference pulse has a random phase than for the case where the phase is known, it may be advantageous to deliberately randomize the phases of the signals in order to reduce the eavesdropper's power. An experimental demonstration of quantum key distribution using active phase randomization was recently reported in \cite{active-phase}.

\nonumsection{Acknowledgments}
\noindent
We thank Daniel Gottesman, Jeff Kimble, Norbert L\"{u}tkenhaus, Bing Qi, Jamin Sheriff, John Sipe,  Kiyoshi Tamaki, and especially Steven van Enk and Masato Koashi for enlightening discussions. We also thank Xiongfeng Ma for help preparing Fig.~\ref{fig:plot}. This work has been supported in part by: the Department of Energy under Grant No. DE-FG03-92-ER40701,  the National Science Foundation under Grant No. PHY-0456720, the National Security Agency (NSA) under Army Research Office (ARO) Contract No. W911NF-05-1-0294, NSERC, Canada Research Chairs Program, Canadian Foundation for Innovation, Ontario Innovation Trust, Premier's Research Excellence Award, Canadian Institute for Photonics Innovations, Canadian Institute for Advanced Research, MITACS, and the Perimeter Institute. Research at the Perimeter Institute is supported in part by the government of Canada through NSERC and by the province of Ontario through MEDT. H.-K. Lo gratefully acknowledges the hospitality of the Institute for Quantum Information at Caltech, where some of this work was done.

\nonumsection{References}

\appendix{\ Bounding the phase error rate}
\label{app:uncertainty}

The argument in Sec.~\ref{sec:security} invokes the uncertainty principle to establish that private key can be extracted from sifted key at the asymptotic rate $R$ shown in eq.~(\ref{key-rate}); this rate depends on an observed bit error rate $\delta_x$ and an unobserved ``phase error rate'' $\delta_y'$ for the signals sent in the $x$ basis. In this Appendix, we derive eq.~(\ref{delta-y-prime}), which expresses $\delta_y'$ in terms of the observed $y$-basis bit error rate $\delta_y$ and the imbalance $\Delta'$ of the quantum coin that determines the basis choice. The derivation is nearly identical to an argument used by Koashi in \cite{koashi}; we include it in this paper so that our security analysis will be self-contained.

The derivation uses the ``Bloch sphere bound'' proved in \cite{koashi-b92}. Note that for an arbitrary state of a single qubit, expectation values of the Pauli operators $X$ and $Z$ obey the inequality
\begin{equation}
\langle X\rangle^2 + \langle Z\rangle^2 \le 1~; 
\end{equation}
the polarization vector in the qubit lies within the Bloch sphere of unit radius. The Bloch sphere bound asserts that a similar inequality applies with high probability if $\langle X\rangle$ and $\langle Z\rangle$ are estimated by randomly sampling from a correlated state of many qubits. That is, consider an arbitrary state of $2n$ qubits, and randomly select a subset of $n$ qubits. Measure in the $x$ basis each of the $n$ qubits in the subset, and let $n\gamma_x$ denote the number of those qubits for which the outcome of the measurement is $X=-1$.  Then measure the rest of the qubits in the $z$ basis, and let $n\gamma_z$ denote the number of qubits for which the outcome is $Z=-1$. Then for any $\varepsilon >0$ and with probability exponentially (in $n$) close to 1, 
\begin{equation}
(1- 2\gamma_x)^2 + (1-2\gamma_z)^2 \le 1 + \varepsilon~.
\end{equation}
Thus, despite the possible correlations among the qubits, the results are constrained by the naive Bloch sphere bound in the limit of large $n$. It is convenient to rewrite this inequality (suppressing the $\varepsilon$) as
\begin{equation}
1-2\gamma_x \le 2\sqrt{\gamma_z(1-\gamma_z)}~.
\end{equation}

We could apply the Bloch sphere bound to the quantum coin; the coin is measured in the $z$ basis to determine whether Alice will encode her key bit in the $x$ basis or the $y$ basis. Thus $\gamma_z$ is the probability that Alice declares the $y$ basis, $1-\gamma_z$ is the probability that Alice declares the $x$ basis, and the ``imbalance'' of the coin is $\Delta'=\gamma_x$. 

However, our goal is to relate the bit error rate $\delta_y$ that is observed when the $y$ basis is declared to the bit error rate $\delta_y'$ that would have been observed when the $x$ basis is declared, if both Alice and Bob had measured in the ``wrong'' basis (the $y$ basis rather than the $x$ basis). Therefore, we will imagine that Alice and Bob always measure in the $y$ basis, and we will divide the signals into two sets --- the set for which Alice and Bob find key bits that agree (the set ``$n$'' for ``no error'') and the set for which Alice and Bob disagree (the set ``$e$'' for ``error''). 

Let $X_{\rm coin}$ denote the Pauli operator $X$ acting on the coin, let $Z_{\rm coin}$ denote the Pauli operator $Z$ acting on the coin, and define
\begin{eqnarray}
\gamma_x^{(e)} &=& {\rm Prob}(X_{\rm coin}=-1 ~|~ Y~{\rm error})~,\nonumber\\
\gamma_z^{(e)} &=& {\rm Prob}(Z_{\rm coin}=-1 ~|~ Y~{\rm error})~,\nonumber\\
\gamma_x^{(n)}&=& {\rm Prob}(X_{\rm coin}=-1 ~|~ {\rm no}~Y~{\rm error})~,\nonumber\\
\gamma_z^{(n)}&=& {\rm Prob}(Z_{\rm coin}=-1 ~|~ {\rm no}~Y~{\rm error})~.
\end{eqnarray}
Applying the Bloch sphere bound to the signals with a $y$-basis error we obtain
\begin{equation}
\label{bloch-bound-error}
1-2\gamma_x^{(e)} \le 2\sqrt{\gamma_z^{(e)}(1-\gamma_z^{(e)})}~,
\end{equation}
and applying it to the signals with no $y$-basis error we obtain
\begin{equation}
\label{bloch-bound-no-error}
1-2\gamma_x^{(n)} \le 2\sqrt{\gamma_z^{(n)}(1-\gamma_z^{(n)})}~.
\end{equation}

Using Bayes's rule, we have
\begin{eqnarray}
&&{\rm Prob}(Z_{\rm coin}=-1, Y~{\rm error})\nonumber\\
&&={\rm Prob}(Z_{\rm coin}=-1)\cdot{\rm Prob}(Y ~{\rm error}~|~Z_{\rm coin}=-1)\nonumber\\
&&= \gamma_z\delta_y\nonumber\\
&&= {\rm Prob}(Z_{\rm coin}=-1~|~Y~{\rm error})\cdot {\rm Prob}(Y~{\rm error})\nonumber\\
&&= \gamma_z^{(e)}\cdot {\rm Prob}(Y~{\rm error})~,
\end{eqnarray}
and likewise
\begin{eqnarray}
&&{\rm Prob}(Z_{\rm coin}=1, Y~{\rm error})\nonumber\\
&&={\rm Prob}(Z_{\rm coin}=1)\cdot{\rm Prob}(Y ~{\rm error}~|~Z_{\rm coin}=1)\nonumber\\
&&= (1-\gamma_z)\delta_y'\nonumber\\
&&= {\rm Prob}(Z_{\rm coin}=1~|~Y~{\rm error})\cdot {\rm Prob}(Y~{\rm error})\nonumber\\
&&= (1-\gamma_z^{(e)})\cdot {\rm Prob}(Y~{\rm error})~.
\end{eqnarray}
Furthermore,
\begin{eqnarray}
&&{\rm Prob}(X_{\rm coin}=-1, Y~{\rm error})\nonumber\\
&&= {\rm Prob}(X_{\rm coin}=-1~|~Y~{\rm error})\cdot {\rm Prob}(Y~{\rm error})\nonumber\\
&&= \gamma_x^{(e)}\cdot {\rm Prob}(Y~{\rm error})~;
\end{eqnarray}
therefore, multiplying both sides of eq.~(\ref{bloch-bound-error}) by ${\rm Prob}(Y~{\rm error})$, we find
\begin{eqnarray}
\label{error-inequality}
&&{\rm Prob}(Y~{\rm error}) - 2~{\rm Prob}(X_{\rm coin}=-1, Y~{\rm error})\nonumber\\
&& \le 2\sqrt{\gamma_z(1-\gamma_z)\delta_y\delta_y'}~.
\end{eqnarray}
By applying the same reasoning to the signals with no $y$-basis error, we find from eq.~(\ref{bloch-bound-no-error}) that
\begin{eqnarray}
\label{no-error-inequality}
&&{\rm Prob}({\rm no}~Y~{\rm error}) - 2~{\rm Prob}(X_{\rm coin}=-1, ~{\rm no}~Y~{\rm error})\nonumber\\
&& \le 2\sqrt{\gamma_z(1-\gamma_z)(1-\delta_y)(1-\delta_y')}~.
\end{eqnarray}
Now adding together eq.~(\ref{error-inequality}) and eq.~(\ref{no-error-inequality}), we have
\begin{eqnarray}
\label{delta-prime-bound}
&&1-2\Delta' =1-2~{\rm Prob}(X_{\rm coin}=-1) \nonumber\\
&\le& 2\sqrt{\gamma_z(1-\gamma_z)}\Big(\sqrt{\delta_y\delta_y'}+\sqrt{(1-\delta_y)(1-\delta_y')}~\Big)\nonumber\\
&\le& \sqrt{\delta_y\delta_y'}+\sqrt{(1-\delta_y)(1-\delta_y')}~.
\end{eqnarray}
This inequality says that $\delta_y'$ must be close to $\delta_y$ if $\Delta'$ is small. 
After some algebra (and after reinstating $\varepsilon$), we may re-express eq.~(\ref{delta-prime-bound}) in the form eq.~(\ref{delta-y-prime}).

In our analysis, we have allowed a basis asymmetry in the detected signals ($\gamma_z\ne 1/2$). This asymmetry might occur because the source leaks a small amount of information about the basis choice, and Eve could use that basis information to (say) block more $x$-basis signals than $y$-basis signals. However, using the available basis information to enhance the basis asymmetry is not a good strategy for Eve. Rather, her attack is most effective when $\delta_y'-\delta_y$ is largest, and eq.~(\ref{delta-prime-bound}) indicates that the maximal possible value of $\delta_y'-\delta_y$ occurs when $x$-basis and $y$-basis signals are detected with equal likelihood ($\gamma_z=1/2$).

In the end, the inequality eq.~(\ref{delta-prime-bound}) has a simple form and in fact it admits a simple interpretation. Recall that if $\rho_x$ and $\rho_y$ are two density operators and the operators $\{E_a\}$ are the elements of a POVM, then there is an inequality relating the fidelity of the two density operators to the statistical overlap of the measurement outcomes:
\begin{equation}
\label{fidelity-inequality}
\sqrt{F(\rho_x,\rho_y)}\equiv \|\sqrt{\rho_x}\sqrt{\rho_y}\|_{\rm tr} \le \sum_a \sqrt{{\rm tr}(\rho_xE_a){\rm tr}(\rho_y E_a)}~.
\end{equation}
Ignore for the moment the correlations among the signal states that might be induced by Eve's attack, and suppose that, for each signal, the joint state of the qubit measured by Alice and of the bosonic modes measured by Bob is $\rho_x$ when Alice declares the $x$ basis and $\rho_y$ when Alice declares the $y$ basis. Recall that for the detector model formulated in Sec.~\ref{sec:modeling}, the probability that Bob's measurement yields a conclusive outcome does not depend on whether Bob measures in the $x$ basis or the $y$ basis; let us denote this probability that Bob's measurement is conclusive as $Q_x$ in the state $\rho_x$ and $Q_y$ in the state $\rho_y$. Let $\{E_a\}$ be the three-outcome POVM that outputs ``inconclusive'' when Bob's measurement is inconclusive, outputs ``agree'' when Bob's $y$ measurement is conclusive and agrees with Alice's $y$ measurement, and outputs ``disagree'' when Bob's $y$ measurement is conclusive and disagrees with Alice's $y$ measurement. Then eq.~(\ref{fidelity-inequality}) becomes
\begin{equation}
\label{fidelity-three-outcome}
\sqrt{F(\rho_x,\rho_y)}~\le~  \sqrt{(1-Q_x)(1-Q_y)} + \sqrt{Q_xQ_y}\left(\sqrt{\delta_y\delta_y'}+ \sqrt{(1-\delta_y)(1- \delta_y')}~\right)~.
\end{equation}
For a fixed value of $Q_x+Q_y$, the right-hand side of eq.~(\ref{fidelity-three-outcome}) is maximized for $Q_x=Q_y$; therefore, defining $Q=(Q_x+Q_y)/2$, we find
\begin{equation}
\sqrt{F(\rho_x,\rho_y)}~\le ~ 1-Q + Q\left(\sqrt{\delta_y\delta_y'}+ \sqrt{(1-\delta_y)(1- \delta_y')}~\right)~.
\end{equation}
But Eve's attack cannot improve the distinguishability of $\rho_x$ and $\rho_y$; thus 
\begin{equation}
1-2\Delta = |\langle \Psi_y|\Psi_x\rangle|\le \sqrt{F(\rho_x,\rho_y)}~,
\end{equation}
and so we obtain
\begin{equation}
1-2\Delta' \le \sqrt{\delta_y\delta_y'}+ \sqrt{(1-\delta_y)(1- \delta_y')}~
\end{equation}
where $\Delta'=\Delta/Q$, in agreement with eq.~(\ref{delta-prime-bound}).

This fidelity argument establishes security against individual attacks, but we have not seen how to turn it into a rigorous proof of security against collective attacks. Recall that we need eq.~(\ref{delta-prime-bound}) to hold with high probability even if (after Eve's attack) the signals are highly correlated with one another. For this purpose it would be natural to use the quantum de Finetti theorem \cite{definetti,definetti2} to show that the state of all the signals can be well approximated by a convex combination of product states. But unfortunately, because the Hilbert space of the signal and reference modes is infinite dimensional, the de Finetti theorem does not provide a useful approximation. Fortunately, though, the argument based on the imbalance of a quantum coin applies for any two-outcome POVM that Bob might apply (after the filtering), irrespective of the dimension of Bob's Hilbert space.

\newpage
\appendix{\ Unambiguous key discrimination}
\label{app:ukd}

In this Appendix,  we elaborate on the unambiguous key discrimination attack against the signal set eq.~(\ref{bb84states-lojp}). It is convenient to express the signals in an alternative basis, which is natural from the perspective of polarization encoding: 
\begin{eqnarray} \label{bb84states-old}
\ket{\tilde 0_Z} &=& e^{-\mu/2} \left(\ket{\rm vac} + \sqrt{2}~\alpha \ket{0_Z} +\dots \right)~, \nonumber \\
\ket{\tilde 1_Z} &=& e^{-\mu/2} \left(\ket{\rm vac} + \sqrt{2}~\alpha \ket{1_Z} +\dots \right)~, \nonumber \\
\ket{\tilde 0_X} &=& e^{-\mu/2} \left(\ket{\rm vac} + \sqrt{2}~\alpha \ket{0_X} +\dots \right)~, \nonumber \\
\ket{\tilde 1_X} &=& e^{-\mu/2} \left(\ket{\rm vac} + \sqrt{2}~\alpha \ket{1_X} +\dots \right) ~;
\end{eqnarray}
here $\{|0_Z\rangle,|1_Z\rangle,|0_X\rangle, |1_X\rangle\}$ are the ideal BB84 states encoded in the polarization of a single photon, and the ellipsis indicates the multi-photon contribution. In this basis, the phase of each signal is $e^{i\theta}=1$, so that $\alpha=\sqrt{\mu/2}$ is real and positive. We assume that $\alpha$ is small, so that multi-photons are unlikely. 
Ignoring the small multi-photon component, the signals reside in a qutrit Hilbert space with basis $\{|{\rm vac}\rangle,|0_Z\rangle,|1_Z\rangle\}$; expanded in this basis, they may be re-expressed as
\begin{eqnarray} \label{bb84vectors}
|\tilde 0_Z\rangle& \approx & e^{-\mu/2}\left( 1, \sqrt{2}~\alpha , 0 \right)  ~,\nonumber \\
|\tilde 0_X\rangle & \approx & e^{-\mu/2}\left( 1, ~{ \alpha } , ~{ \alpha  } \right)  ~,\nonumber \\
|\tilde 1_Z\rangle& \approx & e^{-\mu/2} \left( 1,0, \sqrt{2}~\alpha \right)  ~,\nonumber \\
|\tilde 1_X\rangle & \approx & e^{-\mu/2} \left( 1, ~{ \alpha  } , -{ \alpha  } \right) ~.
\end{eqnarray}

We will describe an intercept/resend attack on BB84 using these signals, based on {\em unambiguous key discrimination}. By exploiting her knowledge of the phase of the signals, Eve performs a POVM with three outcomes: 0, 1, and DK (don't know). The DK outcome is inconclusive, but if either of the other outcomes occurs, then Eve knows with certainty the key bit (0 or 1) encoded in the BB84 signal emitted by the source, though she does not gain any information about which of the two possible BB84 states compatible with that key bit was emitted. Eve blocks the signals when her outcome is DK, but if the outcome is conclusive she sends on to Bob a uniform coherent superposition of the two compatible BB84 states. This procedure generates a bit error rate $\delta = \frac{1}{2} - \frac{1}{2\sqrt{2}}\approx .146$. Evidently, Eve has the same key information as Alice and Bob, so that, if she also knows their protocol for error correction and privacy amplification, she will have perfect knowledge of every bit of the final key. 

The multi-photon component of the state can help Eve, but to keep our analysis simple, we will consider an attack on this source that makes no use of the multi-photons. Eve performs an orthogonal measurement that distinguishes photon number less than two from photon number greater than or equal to two, and she discards the state if the latter outcome is found. (We will be interested in values of $\mu$ that are sufficiently small that Eve would not benefit very much from taking advantage of the multi-photons.) Thus the states she retains are qutrits. She then performs a three-outcome POVM (unambiguous key distribution) to identify the key bit. The two conclusive outcomes of the POVM are projections onto the states:
\begin{eqnarray} \label{conclusive}
|0^\perp\rangle &=& N_{\mu,0} \left( -\sqrt{2}~\alpha -{\alpha } ,~
1+ { 1 \over \sqrt{2}} ,~ {1  \over \sqrt{2}} \right) ~,\nonumber \\
|1^\perp\rangle &=& N_{\mu,1} \left( -{\alpha} ,~
1 +{ 1 \over \sqrt{2}} , ~{1  \over \sqrt{2}} \right) ~,
\end{eqnarray}
where $N_{\mu,0}, N_{\mu,1}$ are normalization factors such that 
\begin{eqnarray}
N_{\mu,0}^{-2}= (2+\sqrt{2})\left[1+\left(\frac{1}{2} +\frac{1}{2\sqrt{2}}\right)\mu\right]~,\nonumber\\
N_{\mu,1}^{-2}= (2+\sqrt{2})\left[1+\left(\frac{1}{2} -\frac{1}{2\sqrt{2}}\right)\mu\right]~.\nonumber\\
\end{eqnarray}
The vector $|0^\perp\rangle$ is orthogonal to both of the two states $|\tilde 0_Z\rangle$ and $|\tilde 0_X\rangle$ that indicate the key bit 0. Hence, if this outcome is found, Eve knows for sure that the key bit could not be 0 and so must be 1. Similarly, the vector $|1^\perp\rangle$ is orthogonal to both of the states $|\tilde 1_Z\rangle$ and $|\tilde 1_X\rangle$ that indicate the key bit 1. 

The vectors $|0^\perp\rangle$ and $|1^\perp\rangle$ are nearly parallel for small $\alpha$. To ensure that all three POVM elements are positive, we may choose 
\begin{eqnarray}
&&E_0= \frac{1}{2}|1^\perp\rangle\langle 1^\perp|~,\quad
E_1=\frac{1}{2}|0^\perp\rangle\langle 0^\perp|~,\nonumber\\
&&E_{\rm DK} = I - E_0 - E_1~.
\end{eqnarray}
(For small positive $\mu$, the strength of the conclusive POVM elements can be pushed up slightly, but this is a small effect that we ignore.) 
Thus we find the probability $p_C$ of a conclusive outcome (taking into account that Eve might detect multi-photons and reject the state)
\begin{eqnarray}
&&\langle \tilde 0_Z|E_0|\tilde 0_Z\rangle = \langle \tilde 0_X|E_0|\tilde 0_X\rangle \nonumber\\
&&= \left(\frac{1}{2} -\frac {1}{2\sqrt{2}}\right) \mu e^{-\mu}\left[1+\left(\frac{1}{2} +\frac{1}{2\sqrt{2}}\right)\mu\right]^{-1}
\end{eqnarray}
if the key bit is 0, and 
\begin{eqnarray}
&&\langle \tilde 1_Z|E_1|\tilde 1_Z\rangle = \langle \tilde 1_X|E_1|\tilde 1_X\rangle \nonumber\\
&&= \left(\frac{1}{2} -\frac {1}{2\sqrt{2}}\right) \mu e^{-\mu}\left[1+\left(\frac{1}{2} -\frac{1}{2\sqrt{2}}\right)\mu\right]^{-1}
\end{eqnarray}
if the key bit is 1. We note that the conclusive outcome is slightly more likely when the key bit is 1. This asymmetry can be traced to the property that the overlap $|\langle\tilde 0_X|\tilde 0_Z\rangle|$ of the two signals that indicate the key bit 0 is slightly larger than the overlap $|\langle\tilde 1_X|\tilde 1_Z\rangle|$ of the two signals that indicate the key bit 1. (For other choices of the phase $e^{i\varphi}$ that determines the plane in the Bloch sphere occupied by the BB84 signals, the asymmetry is substantially larger.) In any case, for either value of the key bit, the probability of a conclusive outcome obeys
\begin{equation}
p_C \ge (.146)\mu e^{-\mu}\left[1+(.854)\mu\right]^{-1}~.
\end{equation}

Now $p_C$ is the probability that Eve resends the signal to Bob, and therefore it is also the probability $p_D$ that Bob detects a signal, if his detector is perfectly efficient. If there were no interference by the eavesdropper (and no loss in the quantum channel connecting Alice and Bob), all non-vacuum signals would be detected, and then $p_D=1-e^{-\mu}=\mu + O(\mu^2)$. If Eve uses the unambiguous key discrimination POVM, then of course $p_C$ vanishes in the limit $\mu\to 0$, but what is noteworthy is that $p_C$ vanishes {\em linearly} with $\mu$. Thus for $\mu$ small, a fraction $\eta\approx .146$ of order one of all the non-vacuum signals sent by Alice are received by Bob. (Since multi-photon signals span a space of even higher dimension, there is a POVM --- unambiguous state discrimination --- that, when conclusive, identifies not just the key bit but also the basis \cite{norbert-usd}; however for that POVM the probability of a conclusive outcome is higher order in $\mu$.)

Having characterized Eve's attack against a source that emits the signal set eq.~(\ref{bb84states-old}), let us now consider the security analysis of a phase-randomized source. We will refer to the source that emits weak coherent states with a definite phase known by the eavesdropper as source P (for phase), and to a  source that emits phase-randomized weak coherent states, with the same mean photon number $\mu$,  as source R (for random). If source P emits the signals eq.~(\ref{bb84states-old}), and Eve launches the intercept/resend attack using unambiguous key discrimination, then Eve has perfect knowledge of every key bit, and Alice and Bob detect a bit error rate of $\delta=.146$. But if source R is used instead, we will see that Alice and Bob can extract private key at a nonzero asymptotic rate for a bit error rate up to $\delta=.189$. For the security analysis of source R, we will adopt a more restricted (and less realistic) model of the detector than the model described in Sec.~\ref{sec:modeling}: the detector applies a basis-independent ``squash'' that maps the incoming signal to a qubit, and then an ideal BB84 measurement is performed on that qubit in the appropriate basis.

We suppose that the source emits a multi-photon signal with probability $p_M$, that Bob detects a fraction $p_D$ of all the signals sent by Alice, and that Eve's attack is unrestricted. Of the signals that are received, the fraction that were emitted as multi-photons is no more than $\Delta=p_M/p_D$; the rest are single photon signals. 

To establish a nonzero key generation rate for a relatively high bit error rate, we will need to consider schemes for key reconciliation and privacy amplification that involve two-way communication between Alice and Bob, for which the argument in \cite{koashi} based on the uncertainty principle does not apply. Instead we can prove security following \cite{ShorPreskill}, by relating the BB84 protocol to a protocol in which the key is generated by measuring noisy entangled pairs shared by Alice and Bob. Private key can be extracted at a positive asymptotic rate if it is possible to distill high fidelity entanglement from the noisy entanglement. Entanglement distillation will succeed if the noisy entangled pairs have a bit error rate and a phase error rate that are both sufficiently small. The bit error rate $\delta$ is inferred directly from the verification test in the BB84 protocol; the phase error rate $\delta_p$ is also inferred, but by a less direct argument.

If the source and detector used in the protocol were perfect, then a symmetry argument would suffice to show $\delta=\delta_p$. This symmetry is broken if the equipment is imperfect, but it is still possible to bound the difference between the two error rates using an appropriate characterization of the imperfections. For the case where Bob has a perfect detector, but Alice's source sometimes emits multi-photons, it can be shown as in eq.~(\ref{delta-p-tagged}) that 
\begin{equation}
|\delta_p-\delta| < \Delta/2~,
\end{equation}
where $\Delta$ is the fraction of all the detected signals that were emitted as multi-photons. Eq.~(\ref{delta-p-tagged}) is an improvement found by Boileau \cite{boileau} of the result $|\delta_p-\delta| =\Delta$ found in \cite{GLLP}. 

The security proof in \cite{ShorPreskill}, which relates one-way privacy amplification to one-way entanglement distillation using quantum error-correcting codes, does not apply for a bit error rate above $\delta=.110$. But security of BB84 was established in \cite{twoway} for a bit error rate as high as $.189$, by relating two-way privacy amplification to entanglement distillation with two-way communication between Alice and Bob. The original argument in \cite{twoway} assumed a perfect source and detector. But the two-way entanglement distillation succeeds if both $\delta$ and $\delta_p$ are below $.189$; therefore the argument can be applied to a protocol with imperfect equipment if there is a strong enough bound on $|\delta-\delta_p|$.

If the bit error rate $\delta$ is .146, then the two-way BB84 protocol is secure for $|\delta-\delta_p| < .189 - .146 = .043$. And for a source that emits phase-randomized coherent states, it suffices that $\Delta < .086$, where 
\begin{eqnarray}
\Delta= \frac{p_M}{p_D} &\le& \left(\frac{\frac{1}{2}\mu^2}{(.146)\mu e^{-\mu}\left[1+(.854)\mu\right]^{-1}}\right)\nonumber\\
&=&(3.42)\mu e^\mu\left[1+(.854)\mu\right]~.
\end{eqnarray}
Thus $\Delta < .086$, and the protocol is provably secure, for $\mu < .0240$. The security proof still applies if Bob's detector, rather than being perfectly efficient, has an efficiency that is independent of the basis in which the detector measures, where whether the detector fires is decided randomly, uninfluenced by the eavesdropper \cite{sheriff}.

We have shown, therefore, that the BB84 QKD protocol is less secure using the phase-coherent source P than using the phase-randomized source R. Eve can exploit her knowledge of the phase of the signals emitted by the source P to implement a POVM that, when its outcome is conclusive, unambiguously identifies the key bit. But for the same bit error rate $\delta\approx.146$, signal strength $\mu$ ($< .0240$), and signal detection rate $p_D\approx .146 \mu$, if the signals have random phases then Alice and Bob can generate a final key about which Eve has negligible knowledge. 

In fact, for source P using the signal set eq.~(\ref{bb84states-old}) we do not have a security proof that establishes {\em any} positive bit error rate $\delta$ such that provably secure key can be generated at a positive asymptotic rate. However, as we have emphasized, it is a different story for the signal set eq.~(\ref{bb84states}). In this paper, we have proven the security of BB84 using these signals, even when Eve knows the phase of the reference pulse.

\end{document}